\documentclass[letterpaper,10pt,conference]{IEEEtran}
\usepackage[cmex10]{amsmath}
\usepackage{array}
\usepackage{tabularx}
\usepackage{graphicx}
\usepackage{epstopdf}
\usepackage{epsfig}
\usepackage{citesort}
\usepackage{amstext}
\usepackage{filecontents}
\usepackage{ragged2e}
\usepackage{float}
\usepackage{caption}
\usepackage[top = 1.1 in, bottom = 0.8 in, left = 1.1 in, right = 1.1 in]{geometry}

\title{
 Sparse Channel Estimation with Gradient-Based Algorithms: A comparative Study
 }

\author{Ahmed M. Abd El-Moaty and  Azzedine Zerguine \\
Electrical Engineering Department, King Fahd University of Petroleum and Minerals\\
Dhahran 31261, Saudi Arabia\\
Email:\{g201307470,azzedine\}@kfupm.edu.sa}
\begin{document}
\maketitle
\thispagestyle{empty}
\pagestyle{empty}

\begin{abstract}
 Channel state information (CSI) is very crucial for any wireless communication systems. Typically, CSI can be characterized at the receiver side using channel impulse response (CIR). Many observations have shown that the CIR of broadband multi path wireless channels are often sparse. To this point, the family of least mean square (LMS)-based algorithms have been widely used to estimate the CIR, unfortunately the performance of LMS family is not much accurate in terms of sparse channel estimation. The Least Mean Mixed Norm (LMMN) algorithm combines the advantages of both the  Least Mean square (LMS) and the Least Mean Fourth (LMF)algorithm, which makes this algorithm stands in a very special position among the family members in terms of  convergence and  steady state error. In this paper, we held a fair comparative study between the LMMN and a number of the LMS-based algorithms, such as the LMS algorithm, the zero-attracting (ZA-LMS) algorithm, and the normalized (NLMS) algorithm. Simulation results  are carried out to compare the performance of all these algorithms with  the LMMN algorithm. The results show that the  LMMN algorithm outperforms the rest of these algorithms in the identification of sparse systems in terms of both fast convergence and the steady state error.
\end{abstract}
\vspace{5 pt}
\begin{IEEEkeywords}
Sparse channels, LMS, Mixed-norm algorithm.
\end{IEEEkeywords}

\IEEEpeerreviewmaketitle

\section{Introduction}\label{int}
\begin{figure}[h]
 \captionsetup{justification=centering}
 \centering
 \includegraphics[scale=0.48]{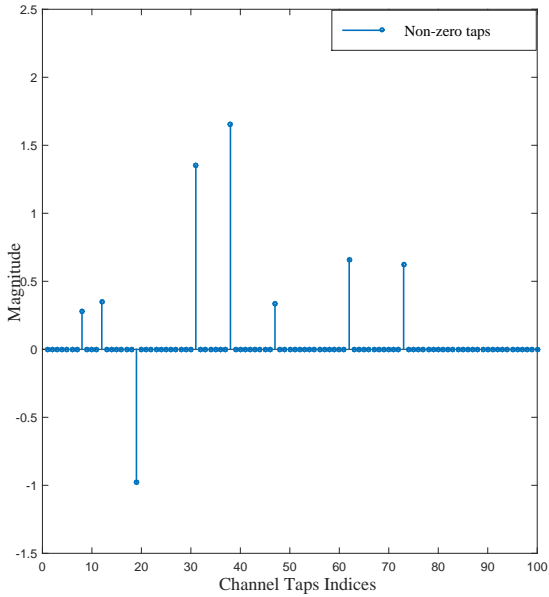}
 \vspace{-3pt}
 \caption{A typical sparse channel with a total length of $100$ and $8$ non-zero taps.}
 \label{fig.1}
 \end{figure}

Wireless communication technologies have  gained a lot of concern in recent decades. One of the important methods to achieve wide bandwidth and high data rate is the broadband transmission, specially in mobile communication systems \cite{adachi2007new}-\cite{cotter2002sparse}. Estimating the coefficients of a communication channel is one of the most dominant challenges since accurate CSI is required for coherent detection at the receiver side. Usually, this can be implemented using adaptive channel estimation (ACE) or adaptive filter algorithms which has been extensively studied in the literature such as the LMS algorithm \cite{sayed2003fundamentals}-\cite{kwong_variable_1992}, the NLMS algorithm \cite{nunoo_variable_2014}, and the LMF algorithm \cite{li2016sparse}. However, most of the classical techniques ignored the fact that most channels in real life are sparse in nature, which means that most of the channel taps are zeros or almost zeros, while a few number of the channel taps are non-zeros \cite{khalifa2013sparse}. Fig.\ref{fig.1} depicts a typical sparse channel. Unfortunately these traditional algorithms was not able to capture the inherent sparsity properties of the sparse broadband multi-path channel.

For the above mentioned reason, the ZA-LMS algorithm \cite{chen_sparse_2009} was introduced. The idea behind the ZA-LMS algorithm  is lying on introducing a penalty term (i.e., $l_1$ norm) in the cost function of the conventional LMS algorithm to achieve fast convergence \cite{salman_two_2013}. However the LMS algorithm suffering from the high sensitivity to the scale of the input data and noise in the poor noise to signal systems. The affine projection methods \cite{li2016low} was introduced to improve the performance of the LMS-based algorithm. The main drawback of these methods is the high computational complexity. The trade-off between the complexity and the performance was solved by introducing the LMS/F algorithm which combines the pros of both LMS and LMF \cite{gui2013least}. Despite its superiority the performance of the LMS/F algorithm was degraded by the effect of the LMF which suffers from the sensitivity to the proximity of the adaptive weights to the optimal Wiener solution. To overcome the sensitivity problem a new  least mean Mixed-Norm (LMMN) algorithm was introduced \cite{chambers1994least}-\cite{zerguine2000convergence}.

In this work, we investigate the performance of the LMMN algorithm with a number of the LMS family to give insights over the channel estimation problem in a wireless communication system using adaptive filtering. Simulation results for different scenarios are assessed to demonstrate the convergence and the steady state performance for the aforementioned algorithms.

This paper is organized as follows. Section II introduces the channel-estimation problem. Section III reviews the adaptive filtering algorithms that have been studied. In section IV we present the simulations performed and discussion. Finally, Section V concludes this work.

\section{Channel Estimation Problem Statement and Formulation}\label{Prob}
Linear time-invariant (LTI) finite impulse response (FIR) is the frequently used system to model communication channels. The output signal can be written in a matrix notation as:
 \begin{align}
\textbf {Y} = \textbf {A} \textbf {w} + \textbf {v},
 \end{align}
where $\textbf {Y}, \textbf {v}  \in R $ are the output signal and noise vectors, respectively. $\textbf A$ is an $N\times M $ Toeplitz convolution matrix and $\textbf {w}$ is the impulse response of an FIR channel of length $M$. The problem of channel-estimation can be seen as that a trial to predict the channel $\textbf {w}$ using a set of training data $\textbf {x}$ and measured outputs $\textbf {Y}$ in the presence of noise $\textbf {v}$. In other words, the objective is to exploit the set of noisy output samples $\textbf {Y}$ and the input data $ \textbf x$ to extract the channel vector $\textbf {w}$. In many of communications system the channels are sparse by nature.

\section{Adaptive Channel Estimation}\label{ACE}
An adaptive filter is mainly a recursive estimator. The main objective of the estimator is finding the optimum set of coefficients of the FIR $\textbf {w}=[w_0,w_1,w_2,......w_{K-1}]^T$ with length $K$ which minimizes a certain cost function, usually the cost function is presented by the error $e(k)$ between a reference signal $d(k)$ and the output of the variable filter $y(k)$, whose coefficients are adjusted progressively. The desired signal can is given by:
\begin{align}
d(k) = \textbf w^T \textbf x(k) + v(k) ,
\end{align}
where $\textbf x(k) = [x(k), x(k-1),...., x(k-K-1)]^T$ and $v(k)$ is a zero-mean additive white Gaussian noise (AWGN) and independent of the input signal. Then the error can be given as:
\begin{align}
e(k) = d(k)-  {\textbf w}^T(k) \textbf x(k)  ,
\end{align}
where ${\textbf w}(k)$ is the estimation of the channel vector at iteration $k$.

There are a large variety of adaptive filtering algorithms to solve the problem of estimation the channel coefficients which are available in the literature \cite{sayed2003fundamentals}. The LMS is perhaps the best known due to its simplicity and ease of implementation. In the ensuing, the LMS algorithm and a variant of gradient-based algorithms are discussed.

\subsection{ The Least Mean-Squares (LMS) Algorithm }
The most famous adaptive filtering algorithms is the Least-Mean Square (LMS) algorithm. The cost function of LMS is defined in terms of the square-error as follows \cite{sayed2003fundamentals}:
\begin{align}
J(k) = \lVert e(k)\rVert^2,
\end{align}
where
 $\lVert e(k)\rVert^2$ is the square of the error. The LMS update equation can be found through derivation By minimizing the above function as:
\begin{align}
\textbf w \normalfont (k) = \textbf w(k-1)+\mu \textbf x^T(k)e(k),
\end{align}
where $\mu$ is the step size which controls the steady state and the convergence of the algorithm.

The pros of the LMS algorithm can be shown as, the algorithm is relatively simple; it requires only a low number of computations per iteration. While the cons of the LMS are presented in dependency of the performance of the LMS on the statistical characteristics of the input signal. Furthermore, the convergence of the LMS algorithm is very slow, which means that a large number of input samples are required for LMS to accurately estimate the channel \cite{sulyman2003convergence}. To deal with this issue the channels considered in this work are mostly populated with zero-valued.

\subsection{The ZA-LMS Algorithm}
The ZA-LMS algorithm is a sparse aware LMS algorithm which adds a penalty term to the original LMS cost function. The added term is used to attract the small nonzero taps to be zeros which matches with the sparsity of the channel. The new cost function with the added $l_{1}$ norm constraint can be shown to:
\begin{align}
J(k) = \lVert e(k)\rVert^2 + \lambda\lVert \textbf w(k)\rVert_1 ,
\end{align}
where $\lambda$ is a regularization parameter used to control the estimation error and the penalty. The update equation through the gradient method can given by:
\begin{align}
\textbf w(k+1) = \textbf w(k) + \mu e(k)\textbf x(k) - \rho f(\textbf w(k)),
\end{align}
where $\rho = \lambda \mu$ and $f(\textbf w(k))$ is the sign function (i.e., $f(\textbf w(k)) = sgn(\textbf w(k))$).

\subsection{The NLMS Algorithm}
One of the drawbacks of the LMS algorithm is the high sensitivity to the scaling of the input. This makes the process of choosing the step-size which guarantees the stability of the algorithm is very hard (if not unfeasible) \cite{sulyman2003convergence}. The NLMS algorithm solves this problem by normalizing the adaptive error update section with the input power. Thus, the NLMS algorithm can be given as:
\begin{align}
\textbf w(k+1) = \textbf w(k)+\mu \frac{e(k)\textbf x(k)}{\varepsilon + \textbf x^T(k)\textbf x(k)},
\end{align}
where $\varepsilon$ is a regulation parameter, which is included in order to avoid large step sizes when $\textbf x(k)\textbf x^T(k)$ becomes small and $\mu$ is the fixed step-size parameter. The variable step size can be used in case of the primary objective of the adaptation (i.e., fast convergence) has been already achieved.

\subsection{The Mixed-Norm LMMN Algorithm}
The Mixed-Norm LMS-LMF algorithm was developed to cover the problem of high steady state errors of NLMS algorithm. The LMMN is a mix between LMS and LMF algorithms. The cost function of LMMN is given by \cite{chambers1994least}:
\begin{align}
J(k) = \alpha E[e^2(k)] + (1-\alpha) E[e^4(k)],
\end{align}
where $\alpha \in [0,1]$ is a positive mixing parameter. If $\alpha = 1$ the algorithm reduces to the conventional LMS algorithm, while if $\alpha = 0$ the algorithm reduces to the LMF algorithm.

In the case of a time varying mixing parameter, the extended variable weight LMMN introduced with the next cost function \cite{zerguine2000convergence}:
\begin{align}
  J(k) = \alpha(k) E[e^2(k)] + (1-\alpha(k)) E[e^4(k)],
\end{align}
and
\begin{align}
\alpha(k+1) = \delta \alpha(k) + \gamma p^2(k) ,
\end{align}
where
\begin{align}
p(k+1) = \beta p(k) + (1-\beta) e(k+1) e(k),
\end{align}
the parameters $\beta$ and $\gamma$ are weighting parameters to control the quality of the estimation of the algorithm both are in the range $[0,1]$ and $\gamma > 0$. It is worth noticing that $\gamma = 0, \delta = 1$ the time varying algorithm is relaxed to the fixed LMMN where $\alpha$ is constant. Consequently, the update equation of the LMMN algorithm is given by
\begin{align}
\textbf w(k+1) = \textbf w(k) + \mu [\alpha (k) e(k) +2 (1- \alpha) e^3 (k)]\textbf x(k) .
\end{align}
\section{Simulations and Discussions} \label{Sim}
For this study, we conducted several simulations to demonstrate the comparison between the different algorithms  mentioned above. We used the Mean Square Deviation (MSD) as a metric to judge the performance of the given algorithms. The MSD is defined by:
\begin{align}
MSD = \sum_{k=1}^{K}{\lvert \textbf w(k) - \hat{\textbf w}(k)\rvert^2} ,
\end{align}
where is $K$ is the length of the data sequence. In this work, over 200 experiments with channel length of 16 taps was used.  $m$ represents the sparsity level and is taking one of the values  $\{ 1,4\}$ and they are normally distributed within the length of the channel. The signal-to-noise ratio is 30 dB. The simulation parameters for the different algorithms are given in Table I and Table II for $m =1$ and $m=4$, respectively.
\begin{table}[h]
 \begin{center}
  \begin{tabular}{||l|p{50mm}||}
 \hline\hline
 Algorithm: & LMS   \\
 \rule{0pt}{3ex}
 Parameters: &   $\mu = 5 \times 10^{-3}$ \\
  \hline\hline
 Algorithm: & ZA-LMS  \\
 \rule{0pt}{3ex}
 Parameters: &    $\mu =  6 \times 10^-3$ , $\rho= 2\times 10^-4 $ \\
  \hline\hline
   Algorithm: & NLMS  \\
   \rule{0pt}{3ex}
   Parameters: &    $\mu = 0.02$ \\
    \hline\hline
   Algorithm: & MN-LMS  \\
   \rule{0pt}{3ex}
  Parameters: &    $\mu = 8 \times 10^-3$ , $\alpha_0 = 0.7$ ,\newline $\gamma = 0.02$ , $ \beta = 0.3$ , $\delta = 0.7 $ \\
        \hline\hline
\end{tabular}
\end{center}
\caption{Simulation Parameters for the different algorithms with sparsity level of 1.}
\end{table}
\begin{table}[h]
 \begin{center}
  \begin{tabular}{||l|p{50mm}||}
 \hline\hline
 Algorithm: & LMS   \\
 \rule{0pt}{3ex}
 Parameters: &   $\mu = 4 \times 10^{-3}$ \\
  \hline\hline
 Algorithm: & ZA-LMS  \\
 \rule{0pt}{3ex}
 Parameters: &    $\mu =  4 \times 10^-3$ , $\rho= 3 \times 10^-5 $ \\
  \hline\hline
   Algorithm: & NLMS  \\
   \rule{0pt}{3ex}
   Parameters: &    $\mu = 0.015$ \\
    \hline\hline
   Algorithm: & MN-LMS  \\
   \rule{0pt}{3ex}
  Parameters: &    $\mu = 4 \times 10^-3$ , $\alpha_0 = 0.85$ ,\newline $\gamma = 0.03$ , $ \beta = 0.9$ , $\delta = 0.95 $ \\
        \hline\hline
\end{tabular}
\end{center}
\caption{Simulation Parameters for the different algorithms with sparsity level of 4.}
\end{table}

\begin{figure}[H]
 \captionsetup{justification=centering}
 \centering
 \includegraphics[scale=0.52]{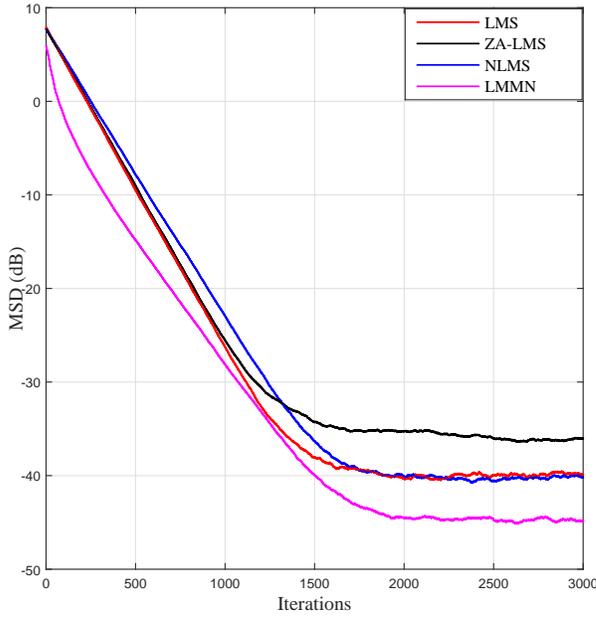}
 \vspace{-3pt}
 \caption{Convergence of different algorithms with m = 1.}
 \label{fig.2}
 \vspace{-0.24cm}
 \end{figure}

\begin{figure}[h]
 \captionsetup{justification=centering}
 \centering
 \includegraphics[scale=0.52]{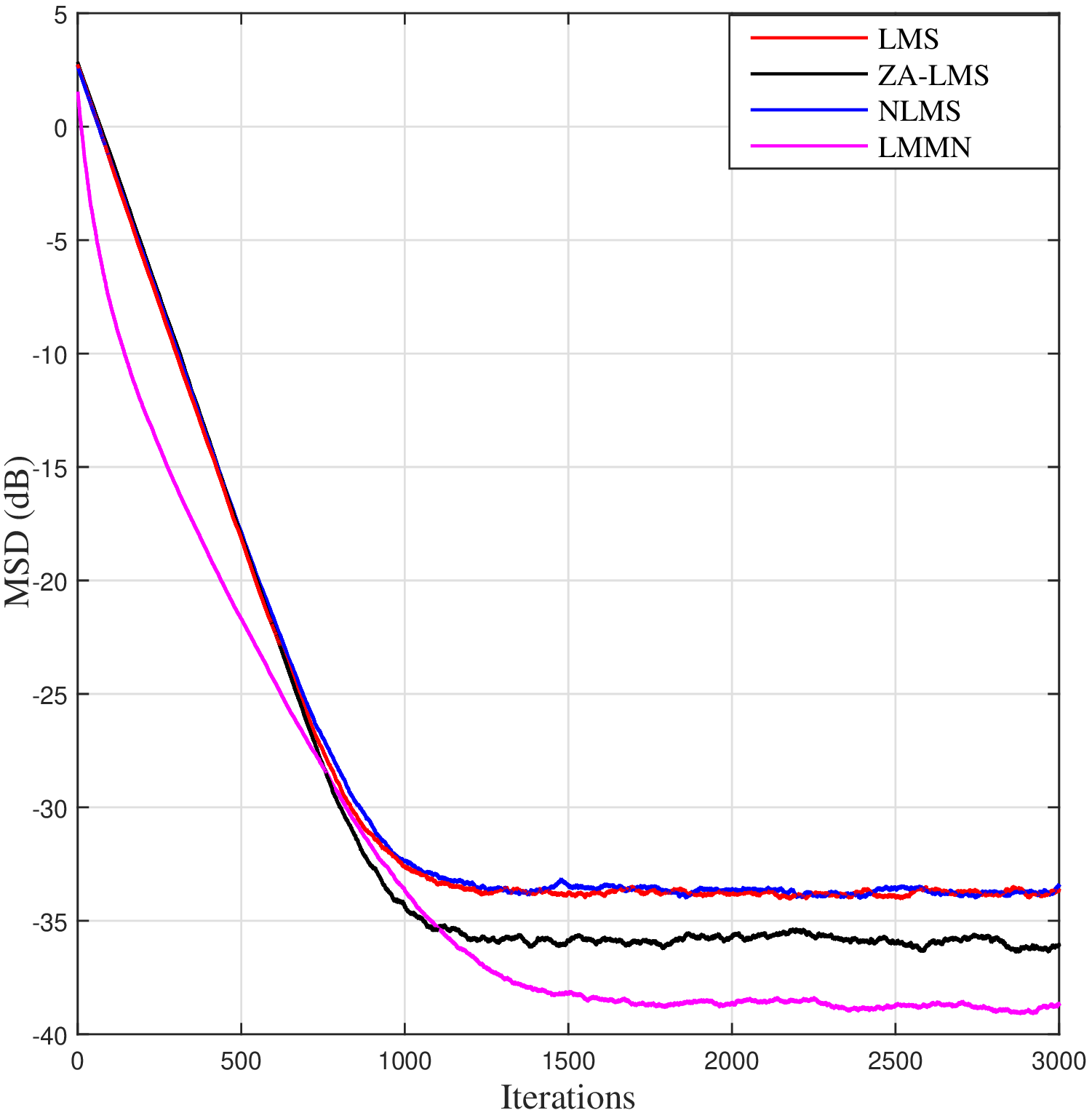}
 \vspace{-3pt}
 \caption{Convergence of different algorithms with m = 4.}
 \label{fig.3}
 \vspace{-0.24cm}
 \end{figure}

\begin{figure}[H]
    \captionsetup{justification=centering}
    \centering
    \includegraphics[scale=0.52]{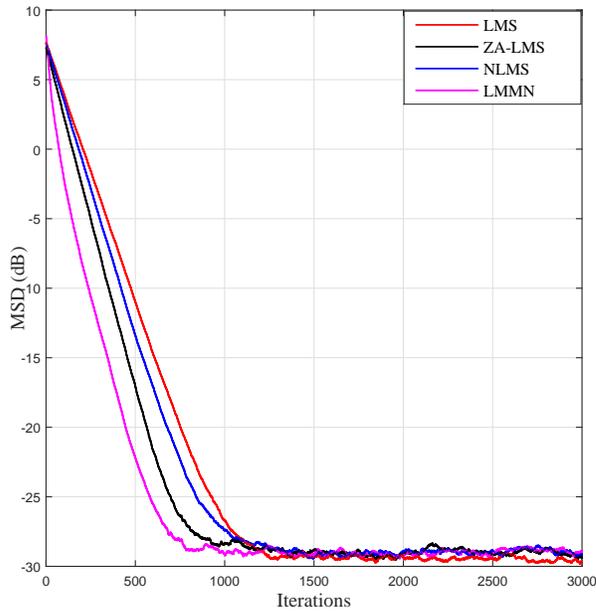}
    \vspace{-3pt}
    \caption{\small{Tracking and steady-state behaviors of 16 tap with $4$ non-zero taps.}}
    \label{fig.4}
    \vspace{-0.24cm}
\end{figure}

Figure \ref{fig.2} shows that for the same speed of convergence, the LMMN algorithm has the lowest steady state error in comparison with the conventional LMS, NLMS, and ZA-LMS algorithms. This is because LMMN algorithm benefits from both features of the LMS and LMF algorithms. Similar behavior is obtained for the case when $m=4$ as depicted in Fig. \ref{fig.3}. In Fig. \ref{fig.4},  it is clear that LMMN has the fastest speed when compared with the other algorithms.

\newgeometry{right= 1.1 in}

 \section{Conclusion} \label{Con}
 In this paper, we conducted a comparative study between different gradient-based algorithms and the combined Mixed-Norm LMS-LMF algorithm in estimation of sparse communication channels. The LMMN was shown as a superior performance over a considered number of the other gradient-based algorithms. The simulation results obtained from the sparse channel estimation were given to show that the LMMN algorithm has fastest convergence and lowest steady-state error  and achieves about 3 dB gain compared to the conventional LMS algorithm when the channel is sparse.
\\

\bf {Acknowledgment:} \normalfont The authors acknowledge the support provided by the Deanship of Scientific Research at KFUPM under Research Grant RG1415.
\\


\restoregeometry
\end{document}